\documentclass[twocolumn,superscriptaddress,prl]{revtex4-1}
\usepackage{mathrsfs,braket}
\usepackage{amssymb, amsbsy, amsmath, latexsym, dsfont, array, layout,
graphicx,mathrsfs,color,ulem,bm}

\newcommand{\one}{\mathds{1}}

\begin{document}

\title{Observation of non-Hermitian bulk-boundary correspondence in quantum dynamics}

\author{Lei Xiao}
\affiliation{Beijing Computational Science Research Center, Beijing 100084, China}
\affiliation{Department of Physics, Southeast University, Nanjing 211189, China}
\author{Tianshu Deng}
\affiliation{CAS Key Laboratory of Quantum Information, University of Science and Technology of China, Hefei 230026, China}
\affiliation{CAS Center For Excellence in Quantum Information and Quantum Physics}
\author{Kunkun Wang}
\affiliation{Beijing Computational Science Research Center, Beijing 100084, China}
\author{Gaoyan Zhu}
\affiliation{Beijing Computational Science Research Center, Beijing 100084, China}
\affiliation{Department of Physics, Southeast University, Nanjing 211189, China}
\author{Zhong Wang}\email{wangzhongemail@tsinghua.edu.cn}
\affiliation{Institute for Advanced Study, Tsinghua University, Beijing, 100084, China}
\author{Wei Yi}\email{wyiz@ustc.edu.cn}
\affiliation{CAS Key Laboratory of Quantum Information, University of Science and Technology of China, Hefei 230026, China}
\affiliation{CAS Center For Excellence in Quantum Information and Quantum Physics}
\author{Peng Xue}\email{gnep.eux@gmail.com}
\affiliation{Beijing Computational Science Research Center, Beijing 100084, China}

\begin{abstract}

{\bf
Bulk-boundary correspondence, a central principle in topological matter relating bulk topological invariants to edge states, breaks down in a generic class of non-Hermitian systems that have so far eluded experimental effort. Here we theoretically predict and experimentally observe non-Hermitian bulk-boundary correspondence, a fundamental generalization of the conventional bulk-boundary correspondence, in discrete-time non-unitary quantum-walk dynamics of single photons. We experimentally demonstrate photon localizations near boundaries even in the absence of topological edge states, thus confirming the non-Hermitian skin effect.
Facilitated by our experimental scheme of edge-state reconstruction, we directly measure topological edge states, which match excellently with non-Bloch topological invariants calculated from localized bulk-state wave functions. Our work unequivocally establishes the non-Hermitian bulk-boundary correspondence as a general principle underlying non-Hermitian topological systems, and paves the way for a complete understanding of topological matter in open systems.
}
\end{abstract}

\maketitle

Topological phases exhibit remarkable properties due to the presence of robust edge states at boundaries. These topologically protected edge states are related to bulk topological invariants through the principle of bulk-boundary correspondence, which is fundamentally important in topological matter~\cite{HKrmp10,QZrmp11}. Intriguingly, recent theoretical studies have shown that the widely held bulk-boundary correspondence apparently breaks down in a broad class of non-Hermitian topological systems~\cite{Lee,Wang1,Wang2,Kunst18,Murakami,Alvarez2018,thomale19}, where the bulk eigenstates are generally localized near boundaries (dubbed the non-Hermitian skin effect~\cite{Wang1,Kunst18}). To correctly account for topological edge states therein, an essential generalization of the conventional bulk-boundary correspondence must be introduced, where the non-Bloch-wave character of bulk states necessitates a new formulation of bulk topological invariants. With rapid progresses in the experimental implementation of non-Hermiticity in synthetic systems ranging from photonics~\cite{PBKMS15,Weimannnm,Parto2018,Zhen2018,OzawaRMP,Zeunerprl,PTsymm2,pxprl,pxdqpt,pxchern,TIlaser} and acoustics~\cite{Chen2018} to vacancy centers in solids~\cite{Du} and cold atoms~\cite{leluo}, non-Hermitian topological systems have generated intense interests recently~\cite{Rudner2009,Shen,nori2017,Ganainy2018,Uedaprx,Ueda18,LeeJY,ESHK11,chenshupra,Lieu}. However, the fundamentally important non-Hermitian bulk-boundary correspondence and the underlying non-Hermitian skin effect have never been experimentally demonstrated yet in any system.

Here we theoretically characterize and experimentally demonstrate the non-Hermitian skin effect and non-Hermitian bulk-boundary correspondence in discrete-time non-unitary quantum walks of single photons. We implement a novel non-unitary Floquet operator with polarization-dependent photon loss that supports Floquet topological phases protected by chiral symmetry. Under a domain-wall configuration, we demonstrate non-Hermitian skin effect by observing that the walker becomes dynamically localized at the boundary with the onset of non-Hermiticity, regardless of its initial state and even in the absence of topological edge states. To demonstrate the non-Hermitian bulk-boundary correspondence, we theoretically calculate the non-Bloch topological invariants defined in a generalized Brillouin zone~\cite{Wang1}, taking into account the generic deviations of localized bulk states from Bloch waves. For our non-unitary quantum walks, two distinct topological invariants exist, corresponding to edge states with quasienergies $\epsilon=0$ and $\epsilon=\pi$, respectively~\cite{Rudner,AO13}.
As both topological edge states and bulk states are localized, it is challenging to unambiguously demonstrate the presence of edge states following the common practice of measuring localized photon population. To overcome this difficulty, we devise a new detection scheme involving quantum-state reconstruction at each time step, which allows us to differentiate edge states from bulk states and fully resolve the quasienergy as well as the internal degrees of freedom of topological edge states. The experimentally measured topological edge states match excellently with the corresponding non-Bloch topological invariants.
In view of the fundamental importance of bulk-boundary correspondence in the study of topological systems, our results pave the way for future studies of topological effects in non-Hermitian systems, and provide a solid groundwork for engineering topological states in open systems.

\begin{figure*}[tbp]
\includegraphics[width=0.6\textwidth]{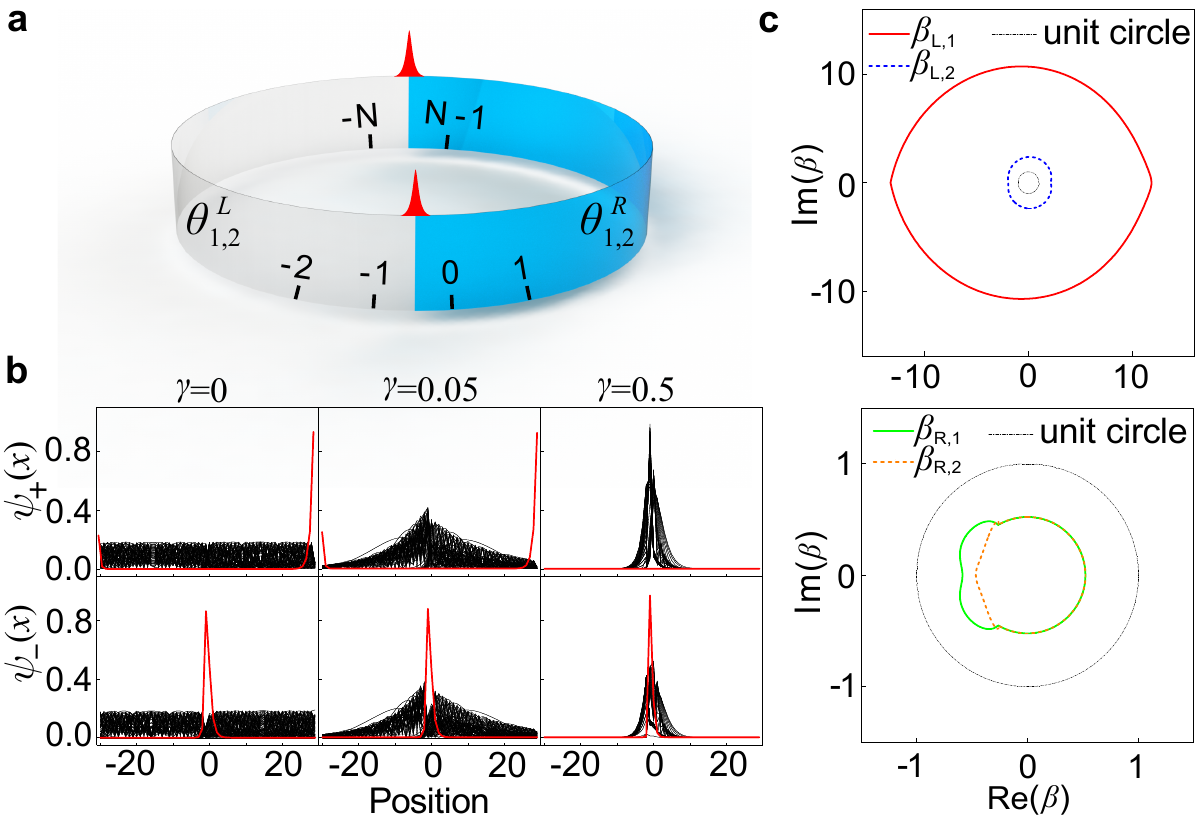}
\caption{{\bf Quantum walks with non-Hermitian skin effect.} {\bf a} Schematic illustration of the domain-wall configuration. The lattice sites are labelled in a cyclic fashion, with the two boundaries located near $x=0$ and $x=-N$ ($x=N-1$).
{\bf b} Spatial distribution of the projected norms of bulk (black) and edge (red) states under various gain-loss parameter $\gamma$, with $\psi_\pm(x)=\big|\left(\langle x|\otimes\langle \pm |\right)|\psi\rangle\big|$. Here
$|\pm\rangle=(|0\rangle\pm|1\rangle)/\sqrt{2}$ are eigenstates of the chiral-symmetry operator, and the bulk-state wave function $|\psi\rangle$ is defined in Eq.~(\ref{eq:bulkwf}).
{\bf c} Generalized Brillouin zones parameterized by the spatial-mode functions $\beta_{\alpha,j}$ for $\gamma=0.5$. Whereas the standard Brillouin zones are indicated by unit circles (solid black line), two distinct generalized Brillouin zones exist for each given bulk.
Parameters for the numerical calculations in {\bf b} and {\bf c} are: $N=30$, $\theta_1^R=0.1875 \pi$, $\theta_1^L=-0.3333 \pi$, $\theta_2^R=0.2 \pi$, and $\theta_2^L=-0.6667 \pi$.
}
\label{fig:fig1}
\end{figure*}

\begin{figure*}
\includegraphics[width=0.8\textwidth]{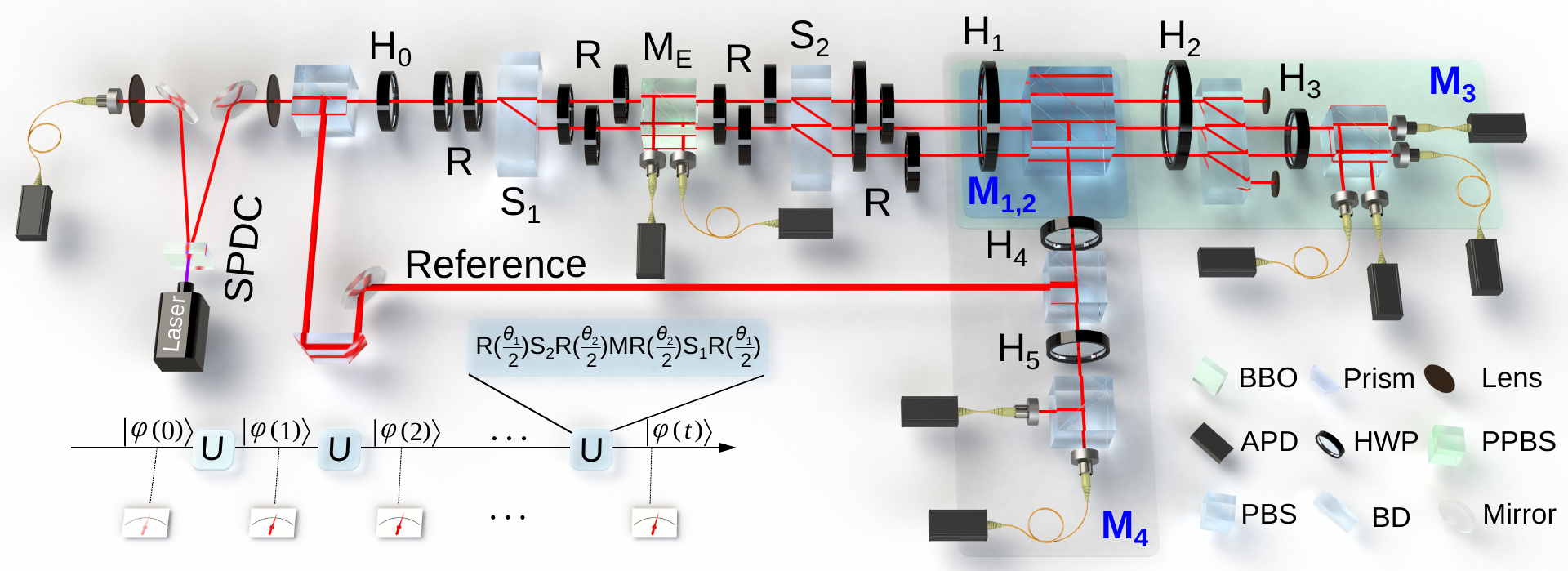}
\caption{{\bf Experimental implementation.}
A pair of photons is created via spontaneous parametric down conversion, of which one serves as a trigger and the other (the walker photon) is projected onto one of the polarization states via a polarizing beam splitter (PBS) and a half-wave plate (HWP) labelled H$_0$. It then proceeds through an interferometric network, composed of HWPs, beam displacers (BDs), and partially polarizing beam splitter (PPBS). For each detection module $M_i$ ($i=1,2,3,4$), the photon is detected by avalanche photodiodes (APDs), in coincidence with the trigger photon. As an illustration, we show the setup for the first step of the quantum-walk configuration as well as a conceptual representation of the multiple-step quantum-walk dynamics (lower-left corner).}
\label{fig:fig2}
\end{figure*}

{\bf Results}

\begin{figure*}[tbp]
\includegraphics[width=\textwidth]{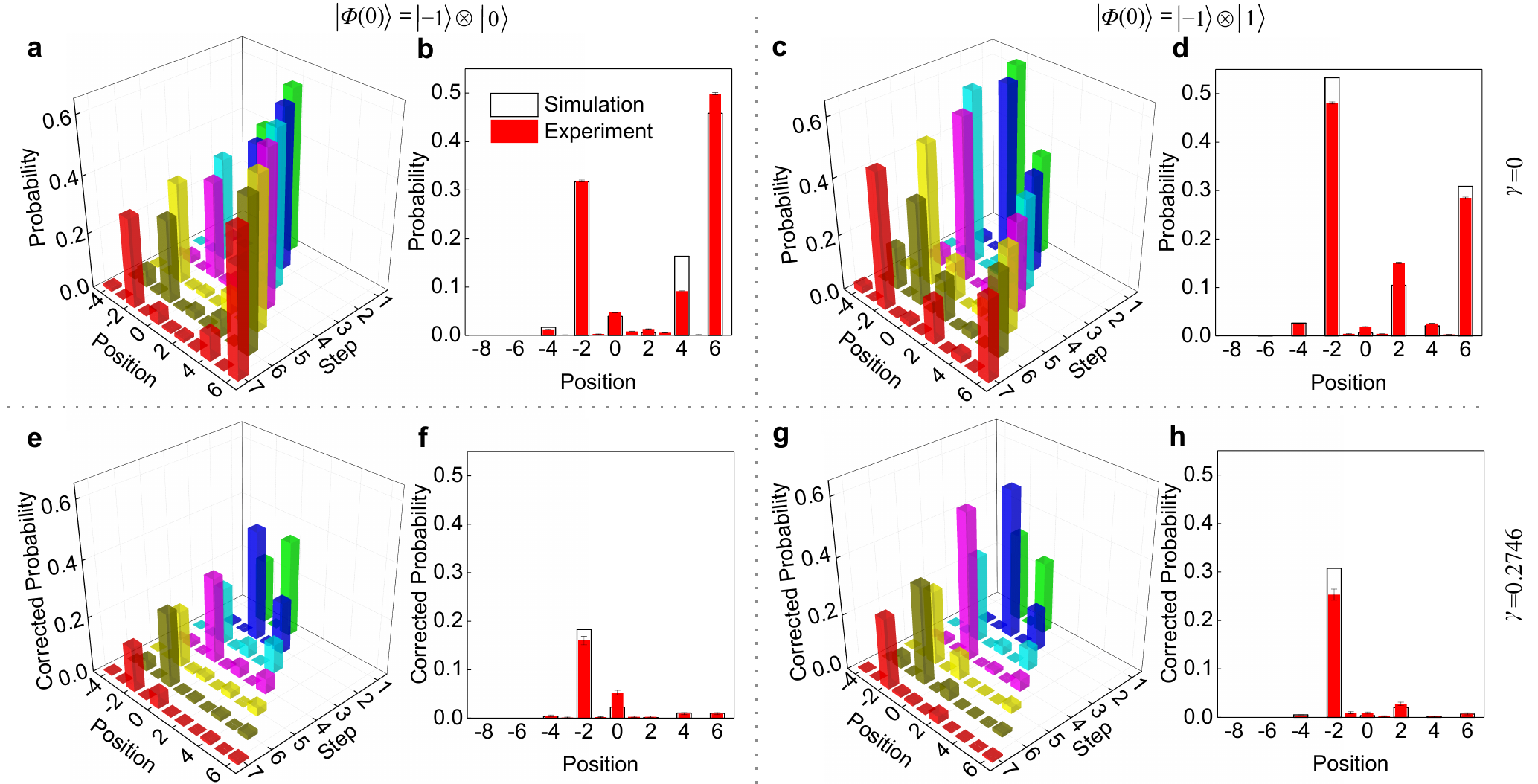}
\caption{{\bf Experimental demonstration of the non-Hermitian skin effect.} (Upper row) Spatial probability distributions of a seven-step unitary quantum walk with $\gamma=0$ and different initial states $|\varPhi(0)\rangle$. Both the time-dependent probability distribution {\bf a, c} and the distribution at the last step {\bf b, d} are shown. (Lower row) Spatial probability distributions of a seven-step non-unitary quantum walk with $\gamma=0.2746$ and different initial states $|\varPhi(0)\rangle$. Both the time-dependent probability distribution {\bf e, g} and the distribution at the last step {\bf f, h} are shown.
The coin parameters are: $\theta_1^R=0.0667 \pi$, $\theta_1^L=0.5625 \pi$, $\theta_2^R=0$, and $\theta_2^L= \pi$.
}
\label{fig:fig3}
\end{figure*}

{\bf Chiral symmetric non-unitary quantum walk.}
We focus on a chiral symmetric non-unitary quantum walk on a one-dimensional lattice governed by the Floquet operator
\begin{align}
U=R(\frac{\theta_{1}}{2})S_{2}R(\frac{\theta_{2}}{2})MR(\frac{\theta_{2}}{2})S_{1}R(\frac{\theta_{1}}{2}),
\label{eq:U}
\end{align}
where the coin operator $R(\theta)$ rotates coin states by $\theta$ about the $y$ axis, and $S_1$ ($S_2$) shifts the walker in the coin state $|1\rangle$ ($|0\rangle$) to the right (left) by one lattice site. Non-unitarity is introduced through $M=\one_w\otimes\left(e^{\gamma}|0\rangle\langle0|+e^{-\gamma}|1\rangle\langle 1|\right)$, where $\one_w$ is the identity operator in lattice modes and $\gamma$ is a tunable gain-loss parameter.

The Floquet operator $U$ has a chiral symmetry with $\sigma_x U\sigma_x=U^{-1}$ ($\sigma_x$ is the standard Pauli matrix),
and the entries of $U$ are real-valued, meaning that its eigenvalues can also be purely real. More importantly, when the lattice has boundaries, $U$ exhibits non-Hermitian skin effect, under which all eigensates of $U$ become localized near the boundaries. This feature distinguishes $U$ from previous experimentally realized non-unitary quantum walks~\cite{PBKMS15,Zeunerprl,PTsymm2,pxprl,pxdqpt,pxchern}. The non-Hermitian skin effect of $U$ is related to that of its effective Hamiltonian defined through $U=e^{-iH_{\rm eff}}$, since quantum-walk dynamics under $U$ can be regarded as a stroboscopic simulation of the non-unitary dynamics driven by the non-Hermitian Hamiltonian $H_{\rm eff}$.

To investigate non-Hermitian skin effect and non-Hermitian bulk-boundary correspondence of $H_{\rm eff}$, we consider a domain-wall configuration on a circle as shown in Fig.~\ref{fig:fig1}a, where the left and right bulks have different coin parameters denoted as $\theta^L_{1(2)}$ and $\theta^R_{1(2)}$, respectively. The lattice sites are indexed in a cyclic fashion, with $x\in J_L$ ($J_{L}=\{x\in \mathbb{Z}|-N\leqslant x\leqslant -1\}$) for the left bulk, and $x\in J_R$ ($J_{R}=\{x\in \mathbb{Z}|0\leqslant x\leqslant N-1\}$) for the right. In Fig.~\ref{fig:fig1}b, we numerically demonstrate that, under such a configuration, both bulk and topological edge states are localized at the boundaries for finite $\gamma$, a hallmark of the non-Hermitian skin effect. Here we identify topological edge states by their quasienergies $\epsilon=0,\pi$ and coin states $|\pm\rangle=(|0\rangle\pm|1\rangle)/\sqrt{2}$.
It follows that the bulk-state wave functions can be written as~\cite{Wang1,tianshu}
\begin{align}
|\psi\rangle=\sum_{\alpha}\sum_{x\in J_\alpha,j}\beta^{x}_{\alpha,j}|x\rangle\otimes|\phi^\alpha_{j}\rangle_c\quad (\alpha=L,R),\label{eq:bulkwf}
\end{align}
where $\beta_{L(R),j}$ is the spatial-mode function for the $j$-th mode in the left (right) bulk, $|\phi^{L(R)}_j\rangle_c$ is the corresponding coin state. For the domain-wall configuration considered here, two spatial modes (labelled by $j=1,2$) exist for each bulk (see Supplemental Information). In the unitary limit with $\gamma=0$, $|\beta_{\alpha,1}|=|\beta_{\alpha,2}|=1$ and we identify $\beta_{\alpha,j}$ as $e^{ik}$, with $k$ being the quasimomentum in the first Brillouin zone and $|\psi\rangle$ reduced to the Bloch wave functions. For a finite $\gamma$ and under the domain-wall configuration, however, $|\beta_{\alpha,j}|\neq 1$, which underlies the non-Hermitian skin effect.

According to the topological band theory, the topological invariants of the Floquet operator $U$ read~\cite{WangPRB,Fruchart}
\begin{align}
\nu_{\epsilon}=\frac{1}{4\pi i}\int_{0}^{2\pi}{\rm Tr}\Big[\sigma_x\Big(\overline{U}_{\epsilon}(k)\Big)^{-1}d\Big(\overline{U}_{\epsilon}(k)\Big)\Big]\quad (\epsilon=0,\pi),
\label{eq:ti}
\end{align}
where $\overline{U}_{\epsilon}(k)$ is the periodized Floquet operator in quasimomentum $k$ space associated with the non-Hermitian effective Hamiltonian $H^{\epsilon}_{\rm eff}=i\ln_{\epsilon} U$, with a branch cut at $\epsilon$ (see Methods). Correspondingly, $\nu_{0(\pi)}$ dictates topological edge state with quasienergy $\epsilon=0$ ($\pi$).

For our domain-wall system with non-Hermitian skin effect, however, these Bloch topological invariants cannot correctly predict topological edge states. The exponential localization of the nominal bulk states at the domain wall suggests that the conventional Brillouin zone with real-valued $k$ should not be the appropriate starting point. Instead, since the Bloch phase factor $e^{ik}$ is replaced by $\beta_{\alpha,j}:=|\beta_{\alpha,j}(p^{\alpha}_j)|e^{ip^{\alpha}_j}$, one should resort to the generalized Brillouin zone~\cite{Wang1}, parameterized by the phase parameter $p_j^\alpha$. As $p_j^\alpha$ varies, the allowed values of $\beta_{\alpha,j}$ form one-dimensional trajectories in the complex plane, with the bulk quasienergy spectrum of the system given by eigenvalues of $H_\text{eff}(e^{ik}\rightarrow \beta_{\alpha,j})$.
Thus, these trajectories (illustrated in Fig.~\ref{fig:fig1}c) fundamentally generalize the concept of Brillouin zones in one-dimensional Hermitian systems to generic one-dimensional non-Hermitian settings, and are identified as the generalized Brillouin zones. The non-Bloch topolgoical invariants $\tilde{\nu}_{0(\pi)}$ are defined over these generalized Brillouin zones (see Methods).

As shown in Fig.~\ref{fig:fig1}c, two distinct generalized Brillouin zones can be found for a single bulk, corresponding to the two distinct spatial modes $j=1$ and $2$, respectively. However, we have numerically checked that
Eq.~(\ref{eq:ti}) yields the same set of non-Bloch topological invariants $\tilde{\nu}_{0(\pi)}$ when integrated over different generalized Brillouin zones of the same bulk (see Supplemental Information).
As we experimentally demonstrate in the following, these non-Bloch topological invariants correctly predict the existence of both types of topological edge states with $\epsilon=0,\pi$, which unambiguously embodies the non-Hermitian bulk-boundary correspondence.

\begin{figure*}
\includegraphics[width=\textwidth]{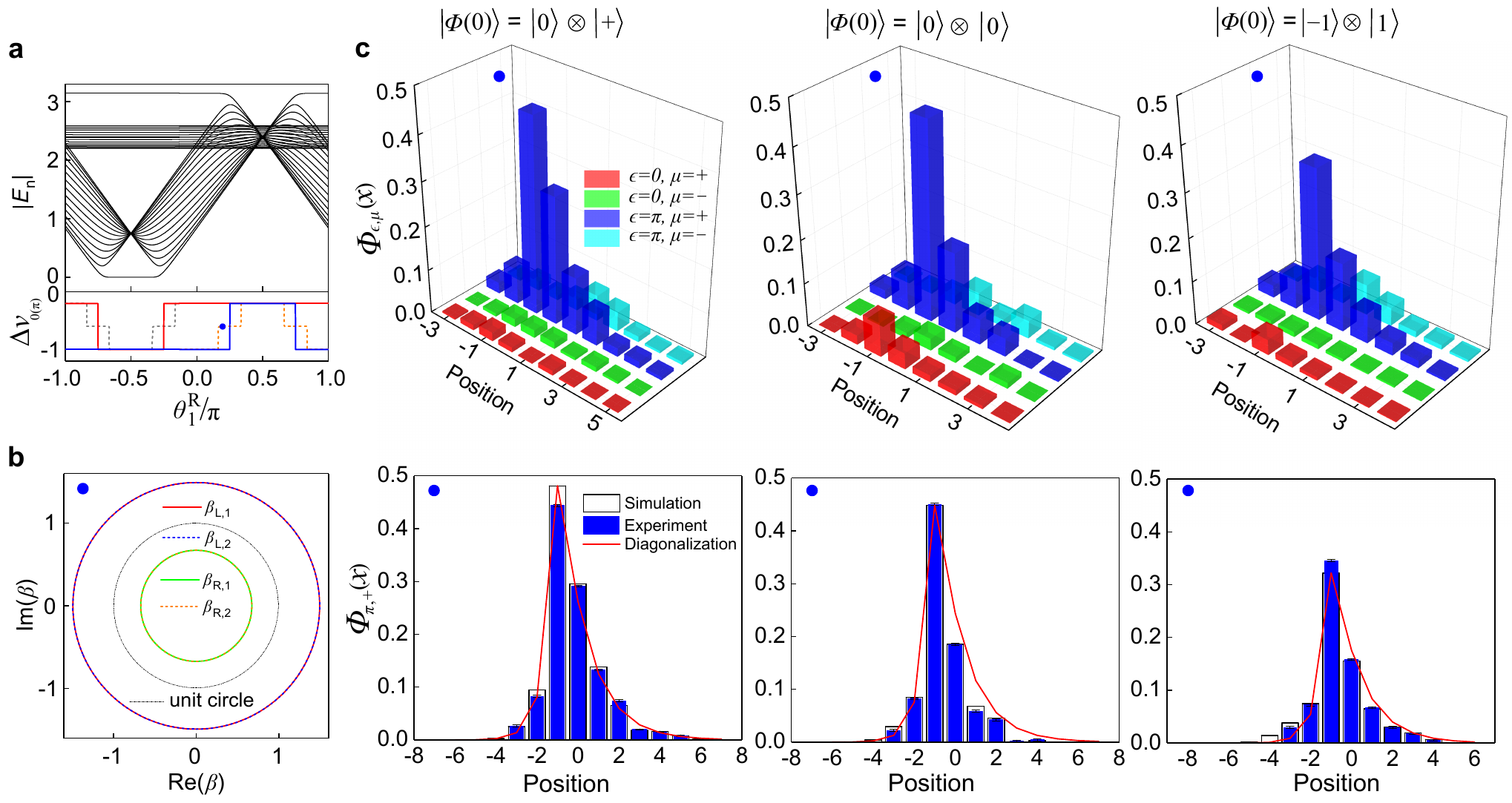}
\caption{{\bf Non-Bloch bulk-boundary correspondence for edge states with $\epsilon=\pi$.} {\bf a} Quasienergy spectrum (black), and winding-number differences for zero- (red) and $\pi$-modes (blue). The non-Bloch winding numbers (solid) are different from the Bloch ones (dashed). The blue dot indicates the parameter for {\bf c}.
{\bf b} Generalized Brillouin zones characterized by $\beta_{\alpha,j}$ on the complex plane.
{\bf c} (Upper row) Experimentally measured $\varPhi_{\epsilon,\mu}(x)$ after the seventh step with different initial states. (Lower row) Comparison between experimentally-measured and numerically-simulated $\varPhi_{\pi,+}(x)$. We also show the scaled norms of the edge state with $\epsilon=\pi$ after the seventh step, calculated by diagonalizing a domain-wall system with $N=15$. Norms of the edge states from diagonalization are scaled to fit the central peak of the numerically-simulated $\varPhi_{\pi,+}(x)$. For all panels, we have $\theta_1^L=0.5625 \pi$, $\theta_2^R= 0.25\pi$, $\theta_2^L=0.75\pi$, and $\gamma=0.2746$. For {\bf b} and {\bf c}, $\theta_1^R=0.18 \pi$.
}
\label{fig:fig4}
\end{figure*}

{\bf Experimental demonstration of non-Hermitian skin effect.}
We experimentally investigate non-unitary quantum-walk dynamics governed by $U$ using a single-photon interferometer setup, as illustrated in Fig.~\ref{fig:fig2}. The coin states are encoded in the photon polarizations, with $|0\rangle$ and $|1\rangle$ corresponding to the horizontally and vertically polarized photons, respectively. The walker states are encoded in the spatial mode of the photons. We experimentally realize a mode-selective loss operator $M_E=\one_w\otimes \left(|0\rangle\langle 0|+\sqrt{1-p}|1\rangle\langle 1|\right)$, which enforces a partial measurement in the basis of $\{|0\rangle,|1\rangle\}$ at every time step. Since $M=e^\gamma M_E$ with $\gamma=-\frac{1}{4}\ln (1-p)$, it is straightforward to map the experimentally implemented dynamics to those under $U$ by multiplying a time-dependent factor $e^{\gamma t}$.

We first demonstrate non-Hermitian skin effect through the local accumulation of population at long times in the absence of topological edge states. We study seven-step quantum-walk dynamics and focus on a single boundary in the domain-wall configuration. Initializing the walker at $x=-1$ but with different coin states, we first implement unitary quantum walks with $\gamma=0$. As illustrated in Fig.~\ref{fig:fig3}a-d, the spatial photon distribution becomes increasingly non-local in time. In contrast, when choosing a finite $\gamma$ (see Fig.~\ref{fig:fig3}e-h), the population distribution after seven steps are still localized near the boundary, regardless of the initial coin state. Crucially, for both the unitary and non-unitary cases, we choose the coin parameters such that no topological edge states exist, which is numerically confirmed by the absence of zero- or $\pi$-modes in the quasienergy spectra. The localization of the population distribution is therefore the unequivocal manifestation of the localization of bulk eigenstates by the non-Hermitian skin effect.

\begin{figure*}
\includegraphics[width=0.75\textwidth]{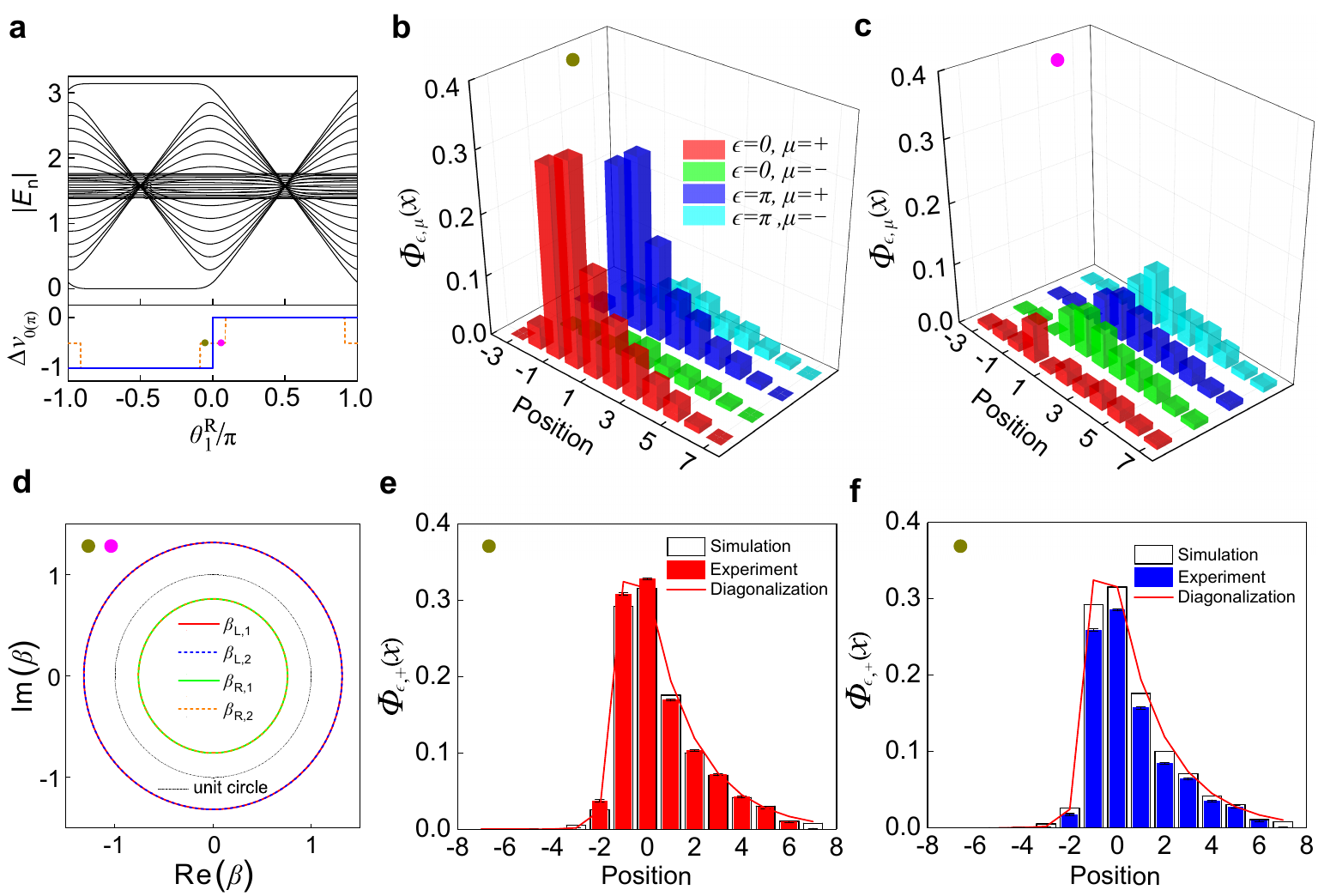}
\caption{{\bf Evidence for non-Bloch bulk-boundary correspondence.} {\bf a} Quasienergy spectrum, and winding-number differences for zero- and $\pi$-modes between the two bulks, with the parameters: $\theta_1^L=0.5625 \pi$, $\theta_2^R=0$, $\theta_2^L= \pi$, and $\gamma=0.2746$. Representations of color and line shapes are the same as those in Fig.~\ref{fig:fig4}. The cyan dot with $ \theta_1^R=-0.0667 \pi$ corresponds to the parameter used in {\bf b, e, f}.
The magenta dot with $\theta_1^R=0.0667 \pi$ corresponds to the parameter used in {\bf c}.
{\bf b, c} Experimentally measured $\varPhi_{\epsilon,\mu}(x)$ for different $\theta_1^R$ after the seventh step, with the initial state $|0\rangle \otimes |+\rangle$.
{\bf d} Generalized Brillouin zones on the complex plane.
{\bf e, f} Comparison between experimentally-measured and numerically-calculated $\varPhi_{\epsilon,+}(x)$ , as well as the scaled norms of the corresponding edge state after the seventh step.
}
\label{fig:fig5}
\end{figure*}

{\bf Non-Hermitian bulk-boundary correspondence.}
We now proceed to a systematic study of non-Hermitian bulk-boundary correspondence by matching the presence of zero- and $\pi$-modes with non-Bloch topological invariants. However, it is challenging to differentiate topological edge states from bulk states by measuring probability distribution, as both states are localized at the boundary under the non-Hermitian skin effect.

To overcome the difficulty, we develop a novel detection scheme which allows us to extract edge-state wave functions with spatial and coin-state resolution. Formally, writing the time-evolved wave function as $|\varPhi(t)\rangle$ (see Methods), we construct the time-integrated wave functions
\begin{align}
|\varPhi_{\epsilon}(t)\rangle=\sum_{t'=0}^t \frac{e^{i\epsilon t'}}{t+1}|\varPhi(t')\rangle\quad (\epsilon=0,\pi).
\label{eq:sumphi}
\end{align}
As we show in the Methods section, in the weighted summation of Eq.~(\ref{eq:sumphi}), time-dependent phases of bulk states cancel out at long times, provided that the quasienergy spectrum underlying the dynamics is purely real. The resulting $|\varPhi_{\epsilon}(t)\rangle$ would converge to the corresponding edge-state wave function with quasienergy $\epsilon$ at sufficiently large time steps, given that the initial state has a finite overlap with the corresponding edge state. Under chiral symmetry, topological edge states are necessarily in the coin state $|\pm\rangle$, which are eigenstates of the symmetry operator. Projecting the integrated wave function Eq.~(\ref{eq:sumphi}) onto the chiral-symmetry basis $|\pm\rangle$, we measure the quantity (see Methods for details)
\begin{align}
\varPhi_{\epsilon,\mu}(x)=\Big|\Big(\langle x|\otimes \langle \mu|\Big)|\varPhi_\epsilon(t)\rangle\Big|\quad (\mu=\pm),
\end{align}
which, as we show in the following, converges to the norm of the corresponding edge-state wave function (apart from a global scaling factor) after seven steps of a quantum walk.
Since a prerequisite for our detection scheme is the presence of purely real energy spectrum, for the following experiments, we choose the coin parameters under which quasienergy spectra of our domain-wall configuration are purely real.

First, we choose the parameters such that a $\pi$-mode edge state exists in the numerically calculated quasienergy spectrum (see Fig.~\ref{fig:fig4}a). Incidentally, the two generalized Brillouin zones in each bulk overlap with each other in this case, as illustrated in Fig.~\ref{fig:fig4}b.
We initialize the walker in three different initial states and let it evolve for seven steps before we measure $\varPhi_{\epsilon,\mu}(x)$. As shown in the upper row of Fig.~\ref{fig:fig4}c, regardless of initial states, a prominent peak exists only for the $\varPhi_{\pi,+}(x)$ measurement, clearly indicating the presence of a $\pi$-mode edge state with the coin state $|+\rangle$. In the lower row of Fig.~\ref{fig:fig4}c, we see that the measured $\varPhi_{\pi,+}(x)$ at the seventh step matches well with the result from numerical simulations for a seven-step quantum walk. Moreover, $\varPhi_{\pi,+}(x)$ approaches the scaled norms of the edge-state wave function, calculated by diagonalizing a finite ($N=15$) domain-wall system under the same parameters. From Fig.~\ref{fig:fig4}a, we see that, for our chosen parameters, only the non-Bloch winding numbers correctly predict the presence of the $\pi$-mode edge state, with $(\Delta\tilde{\nu}_{0},\Delta\tilde{\nu}_{\pi})=(0,-1)$. Here $\Delta\tilde{\nu}_{0(\pi)}=\tilde{\nu}^L_{0(\pi)}-\tilde{\nu}^R_{0(\pi)}$. For comparison, the corresponding Bloch winding numbers are $(\Delta\nu_{0},\Delta\nu_{\pi})=(0,-1/2)$.

We then perform a series of experiments with parameters where: i) both $\epsilon=0$ and $\pi$ edge states exist; ii) no topological state exists.
As shown in Fig.~\ref{fig:fig5}a, in both cases, the theoretically calculated non-Bloch topological invariants correctly predict the existence of topological edge states, whereas the Bloch topological invariants fail to do so. Specifically, in the first case where both types of edge states exist (see Fig.~\ref{fig:fig5}b), $(\Delta\tilde{\nu}_{0},\Delta\tilde{\nu}_{\pi})=(-1,-1)$ for the non-Bloch winding numbers, and $(\Delta\nu_{0},\Delta\nu_{\pi})=(-1/2,-1/2)$ for the Bloch ones. In the second case where there is no edge state (see Fig.~\ref{fig:fig5}c), $(\Delta\tilde{\nu}_{0},\Delta\tilde{\nu}_{\pi})=(0,0)$ for the non-Bloch winding numbers, and $(\Delta\nu_{0},\Delta\nu_{\pi})=(-1/2,-1/2)$ for the Bloch ones. Note that the generalized Brillouin zones for the two cases in Fig.~\ref{fig:fig5}b and~\ref{fig:fig5}c are the same, as shown in Fig.~\ref{fig:fig5}d.
In Figs.~\ref{fig:fig5}e and~\ref{fig:fig5}f, we show the measured $\varPhi_{0,+}(x)$ and $\varPhi_{\pi,+}(x)$ after the last time step, which agree well with results from numerical simulations and with scaled norms of the numerically calculated edge-state wave functions.
Our results thus unambiguously confirm the non-Hermitian bulk-boundary correspondence.

{\bf Discussion.}

We have experimentally established non-Hermitian skin effect and non-Hermitian bulk-boundary correspondence in discrete-time non-unitary quantum-walk dynamics. Our theoretical approach can be extended to treat general non-unitary dynamics, which host rich topological phenomena beyond their unitary counterparts.
Our experiment highlights the versatility of quantum-walk dynamics in studying non-Hermitian topological systems. Specifically, the detection scheme developed in this work allows us to differentiate topological edge states from bulk states localized by the non-Hermitian skin effect, which is valuable for the exploration of non-Hermitian topological systems in general. Overall, our experimental observation of non-Hermitian bulk-boundary correspondence and the theoretical elucidation of its mechanism is of fundamental importance for the understanding of topological phenomena in open systems. From a practical perspective, the new bulk-boundary correspondence established here will be useful in open-system-based topological designs such as topological lasers.

{\bf Methods}

{\bf Experimental implementation.}
As illustrated in Fig.~\ref{fig:fig2}, the walker photon is initialized in the spatial mode $x=-1$ and projected onto one of the polarization states $\ket{\pm}$, $\ket{0}$ or $\ket{1}$ via a PBS and an HWP labelled H$_0$. The coin operator $R$ is implemented by a set of HWPs. The shift operator $S_1$ ($S_2$) is implemented by a BD~\cite{PTsymm2,pxprl,pxdqpt,pxchern}. The mode-selective loss operator $M_E$ is implemented by a PPBS. At each step, after applying $M_E$, photons in the state $\ket{1}$ are reflected by the PPBS with a probability $p$ and the rests continue propagating in the quantum-walk dynamics. The photon is detected by APDs in the detection modules $M_i$ ($i=1,2,3,4$), in coincidence with the trigger photon. Details on these detection modules are given in the Supplemental Information. Photon counts give measured probabilities after correcting for relative efficiencies of the different APDs.

{\bf Non-Bloch topological invariants.}
Bloch topological invariants of the quantum-walk dynamics are calculated through the Fourier component $U(k)$ of the Floquet operator $U$ in Eq.~(\ref{eq:U}). We first define two branches of effective Hamiltonians through $H^\epsilon_{\rm eff}(k)=i\ln_\epsilon U(k)$, where $\epsilon=0,\pi$ indicates the location of the branch cut. It follows that the imaginary part of the function $\ln_\epsilon\lambda$ belongs to the range $\left[\epsilon-2\pi,\epsilon\right)$. We then invoke periodized time evolution operators~\cite{Rudner,WangPRB,Fruchart} to generate topological invariants. For our chiral-symmetric case, we only need its half-period value~\cite{WangPRB,Fruchart}, defined as
\begin{align}
\overline{U}_\epsilon(k)=U_{\frac{1}{2}}(k)e^{\frac{i}{2}H^\epsilon_{\rm eff}},
\end{align}
where the half-period operator $U_{\frac{1}{2}}(k)$ is the Fourier component of $U_{\frac{1}{2}}:=M^{\frac{1}{2}}R(\frac{\theta_{2}}{2})S_{1}R(\frac{\theta_{1}}{2})$, with $M^{\frac{1}{2}}=\begin{pmatrix}e^{\frac{\gamma}{2}}&0\\0&e^{-\frac{\gamma}{2}}\end{pmatrix}$.
The Bloch winding numbers are then given by Eq.~(\ref{eq:ti}) in the main text, where the integration is over the Brillouin zone with $k\in [0,2\pi)$.

To calculate the non-Bloch topological invariants, it is necessary to generalize the Brillouin zone to the complex plane using information from the now localized bulk states. Following the framework outlined in the main text, we replace the standard Bloch phase factor $e^{ik}$ by
$\beta_{\alpha,j}=|\beta_{\alpha,j}(p^{\alpha}_j)|e^{ip^{\alpha}_j}$, whose trajectory forms the $j$-th generalized Brillouin zone of the corresponding bulk. In practice, one needs to make the substitution $k= p^{\alpha}_j-i\ln|\beta_{\alpha,j}(p^\alpha_j)|$ in Eq.~(\ref{eq:ti}) and integrate over the corresponding generalized Brillouin zone. We note that for the domain-wall system here, two generalized Brillouin zones exist for each given bulk, both yielding the same non-Bloch winding numbers.
Details for the calculation of $\beta_{\alpha,j}$
are given in the Supplemental Information.

Finally, we have checked that topological invariants calculated via periodized Floquet operators are the same as those defined with Floquet operators in different time frames~\cite{AO13}, provided that the generalized Brillouin zones are used. As we detail in the Supplemental Information, by defining a shifted Floquet operator
\begin{align}
U'=M^{\frac{1}{2}}R(\frac{\theta_{2}}{2})S_{1}R(\theta_{1})S_{2}R(\frac{\theta_{2}}{2})M^{\frac{1}{2}},
\end{align}
and calculating its winding number $\tilde{\nu}'$ over the generalized Brillouin zone, we have
\begin{align}
\tilde{\nu}_{0(\pi)}=\frac{\tilde{\nu}\pm \tilde{\nu}'}{2}.
\end{align}
Here $\tilde{\nu}$ is the winding number of $U$ calculated over the generalized Brillouin zone.

{\bf Detection scheme for edge states under non-Hermitian skin effect.}
In previous experiments of topological quantum-walk dynamics, topological edge states were detected by observing localization of the probability distribution near the boundary at long times. However, such a practice fails in our system, as both the edge and bulk states are localized. To unambiguously detect edge states in quantum-walk dynamics, we develop a detection scheme based on the fact that quasienergy of edge states is either $\epsilon=0$ or $\epsilon=\pi$, dictated by chiral symmetry. Our scheme relies on reconstructing time-evolved states at each time step. A weighted summation of wave functions at all time steps then selectively retain contribution of edge states with either $\epsilon=0$ or $\epsilon=\pi$, as contributions from all other states cancel out due to interference.

More concretely, the time-dependent wave function can be written as
\begin{equation}
|\varPhi(t)\rangle=U^{t}|\varPhi(0)\rangle=\sum_{n}e^{-iE_{n}t}\varPhi_{n}|\psi_{n}\rangle,\label{eq:psin}
\end{equation}
where $\varPhi_{n}=\langle\chi_{n}|\varPhi(0)\rangle$, $U|\psi_{n}\rangle=e^{-iE_{n}}|\psi_{n}\rangle$, and $\langle\chi_{n}|U^{-1}=\langle\chi_{n}|e^{iE_{n}}$. By definition, $|\psi_n\rangle$ ($\langle\chi_n|$) is the right (left) eigenvector of $U$~\cite{DCB}.

Substituting Eq.~(\ref{eq:psin}) into Eq.~(\ref{eq:sumphi}), we have, for large $t$,
\begin{align}
|\varPhi_{\epsilon}(t)\rangle=\sum_{n}f_{\epsilon}(E_{n})\varPhi_{n}|\psi_{n}\rangle,
\label{eq:sum}
\end{align}
where
\begin{align}
f_{\epsilon}(E_{n})&=\begin{cases}
\frac{1}{t+1}\frac{1-\exp[-i(E_{n}-\epsilon)(t+1)]}{1-\exp[i(E_{n}-\epsilon)]} & E_{n}\neq\epsilon\\
1 & E_{n}=\epsilon
\end{cases},
\end{align}
with $\epsilon=0,\pi$.

When the spectrum $E_{n}$ is completely real, $\underset{t\rightarrow\infty}{{\rm lim}}f_{\epsilon}(E_{n})\rightarrow0$
for $E_{n}\neq 0,\pm\pi$. Only signals from topological edge states would remain in the long-time dynamics, provided that the corresponding coefficient $\varPhi_n$ is non-vanishing, i.e., the initial state has a finite overlap with the left eigenvector of the relevant edge state.

Experimentally, we probe the projection of the wave-function summations $\varPhi_{\epsilon,\mu}(x)=\left|\left(\langle x|\otimes \langle \mu|\right)|\varPhi_\epsilon(t)\rangle\right|$, such that the internal- and external-degrees of freedom of both types of topological edge states ($\epsilon=0,\pi$) are fully resolved.

{\bf Experimental implementation of edge-state detection.}
We probe $\varPhi_{\epsilon,\mu}(x)$ from our experimental reconstruction of $|\varphi(t)\rangle$ at each time step. Here $|\varphi(t)\rangle$ is the experimentally realized time-dependent state (with $M_E$ rather than $M$), which is related to $|\varPhi(t)\rangle$ through $|\varphi(t)\rangle=e^{-\gamma t}|\varPhi(t)\rangle$. It follows that
\begin{align}
\varPhi_{\epsilon,\mu}(x)=\left|(\langle x|\otimes\langle \mu|)\sum_{t'=0}^t\frac{e^{i\epsilon t'}}{t+1}e^{\gamma t'}|\varphi(t')\rangle\right|.\label{eq:signcal}
\end{align}
Since $U$ and initial states are purely real in the basis of $\{|\pm\rangle\}$, we have the expansion
\begin{align}
\ket{\varphi(t)}=\sum_x \left[p_+(t,x)\ket{x}\otimes\ket{+}+p_-(t,x)\ket{x}\otimes\ket{-}\right],\label{eq:phiexp}
\end{align}
where the coefficients $p_\mu(t,x)$ are also real.
Based on these, we perform four distinct measurements $M_i$ ($i=1,\cdots,4$) to reconstruct $|\varphi(t)\rangle$ in the basis $\{|\pm\rangle\}$. This amounts to measuring the absolute values, the relative signs and a global sign of the real coefficients $\{p_\pm(t,x)\}$, as we detail in the following. All measurement modules are shown in Fig.~\ref{fig:fig2}.

First, we measure the absolute values $|p_\pm(t,x)|$. After the $t$-th step, photons in the spatial mode $x$ are sent to a detection unit $M_1$, which consists of an HWP (H$_1$) at $22.5^\circ$, a PBS and APDs. $M_1$ applies a projective measurement of the observable $\sigma_x$ on the polarization of photons. The counts of the horizontally polarized photons $N_H(t,x)$ and vertically polarized ones $N_V(t,x)$ are registered by the coincidences between one of the APDs in the detection unit, and the APD for the trigger photon. The measured probability distributions are $P_{H(V)}(t,x)=N_{H(V)}(t,x)/\sum_{x}\left[N_H(t,x)+N_V(t,x)+\sum_{t'=0}^{t}N_L(t',x)\right]$, with $N_L(t,x)$ the photon loss caused by the partial measurement $M_E$. The square root of the probability distribution $P_{H(V)}(t,x)$ corresponds to $|p_\pm(t,x)|$.

Second, we determine the relative sign between the amplitudes $p_+(t,x)$ and $p_-(t,x)$ via the detection unit $M_2$. The only difference between $M_2$ and $M_1$ is that the setting angle of the HWP (H$_1$) is set to $0$, i.e., a projective measurement of the observable $\sigma_z$ on the polarization of photons. The difference between the probability distributions of the horizontally and vertically polarized photons is given by
\begin{align}
P_H(t,x)-P_V(t,x)=2p_+(t,x)p_-(t,x),
\end{align}
which determines the relative sign between $p_+(t,x)$ and $p_-(t,x)$.

Third, we probe the relative sign between $p_\pm(t,x)$ and $p_\pm(t,x-1)$, which is necessary to calculate the summation of wave functions at each time step.
To this end, we add a detection unit $M_3$ behind $M_{1}$ and $M_2$ (see Fig.~\ref{fig:fig2}), which consists of two HWPs at $22.5^\circ$ (H$_2$ and H$_3$), a BD, a PBS and APDs. We use a BD to combine the horizontally polarized photons in the spatial mode $x$ and the vertically polarized photons in the spatial mode $x-1$. After a projective measurement on the polarizations of photons via H$_3$ and the following PBS, the difference in probability distribution of the resulting photons between the two polarizations is
\begin{align}
P_H(t,x)-P_V(t,x)=p_+(t,x)p_+(t,x-1)
\end{align}
when H$_1$ is set at $22.5^\circ$ in $M_{1(2)}$, and
\begin{align}
P_H(t,x)-P_V(t,x)=p_-(t,x)p_-(t,x-1)
\end{align}
when the setting angle of H$_1$ is at $-22.5^\circ$.

Finally, we determine the global sign of $p_\pm(t,x)$ relative to the reference photons, which are reflected by the PBS used for the preparation of the initial coin state. For this step, we only need to determine the sign of $p_\pm(t,x_w)$ for an arbitrary position $x_w$ at each time step. A natural choice of $x_w$ is the position where the walker and reference photons have comparable counts.
Assuming the reference photons have an amplitude $a$ ($a>0$), we determine the relative sign between the amplitudes of the reference photons and the walker photons at $x_w$ after $t$ steps.
We remove the detection unit $M_3$ and keep the optical elements of $M_{1(2)}$. Only photons reflected by the PBS in $M_{1(2)}$ are relevant here. These photons, after passing through an HWP (H$_4$ at $45^\circ$), are combined with the reference photons at a PBS. A projective measurement is then applied on the polarization of the photons via H$_5$ at $22.5^\circ$ and the last PBS. The difference between the probabilities of the photons with different polarizations is
\begin{align}
P_H(t,x_w)-P_V(t,x_w)=2ap_-(t,x_w),
\end{align}
which determines the global sign of $p_-(t,x_w)$ as $a$ is positive.
Since we have determined the relative sign between $p_+(t,x)$ and $p_-(t,x)$ via $M_2$ and that between $p_\pm(t,x)$ and $p_\pm(t,x-1)$ via $M_3$, the global sign of $p_\pm(t,x)$ is also determined for an arbitrary $x$.

With the above steps, we reconstruct $\ket{\varphi(t)}$ in the basis of $\ket{\pm}$, which enables us to calculate $\varPhi_{\epsilon,\mu}(x)$ according to Eq.~(\ref{eq:signcal}).

{\bf Acknowledgements}
This work has been supported by the Natural Science Foundation of China (Grant No. 11674056, No. 11674189) and the Natural Science Foundation of Jiangsu Province (Grant No. BK20160024). WY acknowledges support from the National Key Research and Development Program of China (Grant Nos. 2016YFA0301700 and 2017YFA0304100).





\clearpage

\begin{widetext}
\renewcommand{\thesection}{\Alph{section}}
\renewcommand{\thefigure}{S\arabic{figure}}
\renewcommand{\thetable}{S\Roman{table}}
\setcounter{figure}{0}
\renewcommand{\theequation}{S\arabic{equation}}
\setcounter{equation}{0}

\section{Supplemental Information for ``Observation of non-Hermitian bulk-boundary correspondence in quantum dynamics''}

In this Supplemental Information, we provide details on calculations of generalized Brillouin zone, non-Bloch topological invariants, and additional supporting experimental data.

\subsection{Bulk-state wave functions, generalized Brillouin zones, and non-Bloch topological invariants}

We consider quantum-walk dynamics governed by the Floquet operator $U=FMG$. Here
\begin{align}
F&=R[\frac{\theta_{1}(x)}{2}]S_{2}R[\frac{\theta_{2}(x)}{2}],\nonumber\\
G&=R[\frac{\theta_{2}(x)}{2}]S_{1}R[\frac{\theta_{1}(x)}{2}],
\end{align}
where the coin-rotation operator $R$ and the shift operator $S$ are given by
\begin{align}
R(\theta)&=\one_w\otimes e^{-i\theta\sigma_{y}},\nonumber\\
S_{1}&=\sum_{x}|x\rangle\langle x|\otimes|0\rangle\langle0|+|x+1\rangle\langle x|\otimes|1\rangle\langle1|,\nonumber\\
S_{2}&=\sum_{x}|x-1\rangle\langle x|\otimes|0\rangle\langle0|+|x\rangle\langle x|\otimes|1\rangle\langle1|.
\end{align}
Here $\one_w=\sum_x|x\rangle\langle x|$, and $-N\leqslant x\leqslant N-1$ is the site index of the lattice.
For a domain-wall configuration on a circle with $2N$ lattice sites (see Fig.~1 of the main text), we adopt a cyclic index such that $|x-1\rangle|_{x=-N}=|N-1\rangle$ and $|x+1\rangle|_{x=N-1}=|-N\rangle$. We also have
\begin{align}
\begin{cases}
\begin{array}{c}
\theta_{1(2)}(x)=\theta_{1(2)}^{L}\\
\theta_{1(2)}(x)=\theta_{1(2)}^{R}
\end{array} & \begin{array}{c}
x\in J_L\\
x\in J_R
\end{array}\end{cases},
\end{align}
where $\theta_{1(2)}^{\alpha}$ represent coin parameters of the left ($\alpha=L$) and right ($\alpha=R$) bulk, $J_{L}=\{x\in \mathbb{Z}|-N\leqslant x\leqslant -1\}$ and $J_{R}=\{x\in \mathbb{Z}|0\leqslant x\leqslant N-1\}$.

We then rewrite $U$ as
\begin{align}
U=\sum_{x}\Big[|x\rangle\langle x+1|\otimes A_{m}(x)+|x\rangle\langle x-1|\otimes A_{p}(x)+|x\rangle\langle x|\otimes A_{s}(x)\Big],
\end{align}
where the site-dependent coin-state operators $A_{m,p,s}(x)$ are given by
\begin{align}
A_{m}(x)&=F_{m}(x+1)MG_{s}(x+1),\nonumber\\
A_{p}(x)&=F_{s}(x)MG_{p}(x-1),\\
A_{s}(x)&=F_{s}(x)G_{s}(x)+F_{m}(x+1)G_{p}(x),\nonumber
\end{align}
with
\begin{align}
F_{m}(x)&=R[\frac{\theta_{1}(x-1)}{2}]P_0R[\frac{\theta_{2}(x)}{2}],\nonumber\\
F_{s}(x)&=R[\frac{\theta_{1}(x)}{2}]P_1R[\frac{\theta_{2}(x)}{2}],\nonumber\\
G_{s}(x)&=R[\frac{\theta_{2}(x)}{2}]P_0R[\frac{\theta_{1}(x)}{2}],\nonumber\\
G_{p}(x)&=R[\frac{\theta_{2}(x+1)}{2}]P_1R[\frac{\theta_{1}(x)}{2}],
\end{align}
and $P_0=|0\rangle\langle0|$, $P_1=|1\rangle\langle1|$.

Following Refs.~\cite{Wang1,tianshu}, we write the general eigenstate of $U$ as $|\psi\rangle=|\psi^R\rangle+|\psi^L\rangle$, with
\begin{align}
|\psi^\alpha\rangle&=\sum_{x\in J_\alpha,j}\beta^{x}_{\alpha,j}|x\rangle\otimes|\phi^\alpha_{j}\rangle_c \quad (\alpha=L,R),
\end{align}
where $|\phi^\alpha\rangle_c$ is the coin state of the corresponding bulk and $\beta_\alpha$ is the spatial-mode function.

From the eigenstate equation \begin{equation} U|\psi\rangle=\lambda|\psi\rangle, \end{equation}
we have
\begin{equation}
\left(A^\alpha_{m}\beta_\alpha+\frac{A^\alpha_{p}}{\beta_\alpha}+A^\alpha_{s}-\lambda\right)|\phi^\alpha\rangle_c=0,
\label{eq:eigenA}
\end{equation}
where $A^\alpha_{m,p,s}$ are the corresponding coin operators in the bulk, with $A^L_{m,p,s}=A_{m,p,s}(x)$ ($-N+1\leqslant x\leqslant -2$) and $A^R_{m,p,s}=A_{m,p,s}(x)$ ($2\leqslant x\leqslant N-2$).

Equation~(\ref{eq:eigenA}) supports non-trivial solutions for
\begin{equation}
\det\left[A_{m}^{\alpha}\beta_{\alpha}+A_{p}^{\alpha}\frac{1}{\beta_{\alpha}}+A_{s}^{\alpha}-\lambda \right]=0.
\label{eq:char}
\end{equation}
Since Eq.~(\ref{eq:char}) is quadratic in $\beta_{\alpha}$, there exist two solutions $\beta_{\alpha,j}$ with $j=1,2$. Correspondingly, eigenstates of the bulk can be written as
\begin{align}
|\psi^\alpha\rangle&=\sum_{x\in J_\alpha,j=1,2}\beta^{x}_{\alpha,j}|x\rangle\otimes|\phi^\alpha_j\rangle_c.
\label{eq:psiLR}
\end{align}

The domain-wall boundary condition is enforced by substituting Eq.~(\ref{eq:psiLR}) into $U|\psi\rangle=\lambda|\psi\rangle$ at the boundaries ($x=-N,-1,0,N-1$). Making use of Eq.~(\ref{eq:eigenA}), we derive a set of linear equations
$M\left[|\phi^L_1\rangle_c,|\phi^L_2\rangle_c,|\phi^R_1\rangle_c,|\phi^R_2\rangle_c\right]^T=0$, where
\begin{align}
&M=\nonumber\\
&\begin{pmatrix}
-A_{p}^{L}\beta_{L,1}^{-N-1} & -A_{p}^{L}\beta_{L,2}^{-N-1} & A_{p}(-N)\beta_{R,1}^{N-1} & A_{p}(-N)\beta_{R,2}^{N-1}\\
A_{p}^{L}\beta_{L,1}^{-2}+[A_{s}(-1)-\beta]\beta_{L,1}^{-1} & A_{p}^{L}\beta_{L,2}^{-2}+[A_{s}(-1)-\lambda]\beta_{L,2}^{-1} & A_{m}(-1) & A_{m}(-1)\\
A_{p}(0)\beta_{L,1}^{-1} & A_{p}(0)\beta_{L,2}^{-1} & -A_{p}^{R}\beta_{R,1}^{-1} & -A_{p}^{R}\beta_{R,2}^{-1}\\
A_{m}(N-1)\beta_{L,1}^{-N} & A_{m}(N-1)\beta_{L,2}^{-N} & A_{p}^{R}\beta_{R,1}^{N-2}+[A_{s}(N-1)-\lambda]\beta_{R,1}^{N-1} & A_{p}^{R}\beta_{R,2}^{N-2}+[A_{s}(N-1)-\lambda]\beta_{R,2}^{N-1}
\end{pmatrix}
\label{eq:M}
\end{align}

Non-trivial solutions exist only when the $8$-by-$8$ coefficient matrix $M$ satisfies $\det (M)=0$ in the thermodynamic limit $N\rightarrow\infty$. Making use of Eq.~(\ref{eq:eigenA}), the condition $\det (M)=0$ is simplified to
\begin{equation}
a_1\frac{1}{\beta_{L,1}^{N}}\beta_{R,1}^{N}+a_2\frac{1}{\beta_{L,2}^{N}}\beta_{R,1}^{N} +a_3\frac{1}{\beta_{L,1}^{N}}\beta_{R,2}^{N}+a_4\frac{1}{\beta_{L,2}^{N}}\beta_{R,2}^{N} +b_{L}\frac{1}{\beta_{L,1}^{N}}\frac{1}{\beta_{L,2}^{N}}+b_{R}\beta_{R,1}^{N}\beta_{R,2}^{N}=0,
\label{eq:eq7}
\end{equation}
where $\{a_1,a_2,a_3,a_4,b_L,b_R\}$ are some coefficients whose exact forms are not important for the following discussion.

To proceed further, we need to sort the following terms
\begin{align}
\Big\{|\frac{\beta_{R,1}}{\beta_{L,1}}|,|\frac{\beta_{R,1}}{\beta_{L,2}}|,
|\frac{\beta_{R,2}}{\beta_{L,1}}|,|\frac{\beta_{R,2}}{\beta_{L,2}}|,
|\frac{1}{\beta_{L,1}\beta_{L,2}}|,|\beta_{R,1}\beta_{R,2}|\Big\}.
\end{align}
This is because in the thermodynamic limit ($N\rightarrow\infty$), only terms with the largest absolute values survive in Eq.~(\ref{eq:eq7}).
Without loss of generality, we take $|\beta_{\alpha,1}|\geqslant |\beta_{\alpha,2}|$ and discuss the order of these terms case by case.
For example, when
$|\beta_{L,2}\beta_{R,2}|\leqslant1\,\,\text{and}\,\,|\beta_{L,2}\beta_{R,1}|\leqslant1\,\,\text{and}\,\,|\beta_{L,1}\beta_{R,2}|\leqslant1$,
the largest two terms are $|\frac{1}{\beta_{L,1}\beta_{L,2}}|$ and $|\frac{\beta_{R,1}}{\beta_{L,2}}|$. Eq.~(\ref{eq:eq7}) is then reduced to
\begin{align}
a_2\frac{1}{\beta_{L,2}^{N}}\beta_{R,1}^{N}+b_{L}\frac{1}{\beta_{L,1}^{N}}\frac{1}{\beta_{L,2}^{N}}=0.
\end{align}
It follows that, in the thermodynamic limit, $|\beta_{L,1}\beta_{R,1}|=1$.

Exhausting all the possible scenarios, we rewrite Eq.~(\ref{eq:eq7}) as
\begin{equation}
\zeta(\beta_{\alpha,j})=0,\label{eq:zeta}
\end{equation}
where
\begin{equation}
\zeta(\beta_{\alpha,j}):=\begin{cases}
|\beta_{L,1}\beta_{R,1}|-1, & |\beta_{L,2}\beta_{R,2}|\leqslant1\,\,\text{and}\,\,|\beta_{L,2}\beta_{R,1}|\leqslant1\,\,\text{and}\,\,|\beta_{L,1}\beta_{R,2}|\leqslant1,\\
|\beta_{L,2}\beta_{R,2}|-1, & |\beta_{L,1}\beta_{R,1}|\geqslant1\,\,\text{and}\,\,|\beta_{L,1}\beta_{R,2}|\geqslant1\,\,\text{and}\,\,|\beta_{L,2}\beta_{R,1}|\geqslant1,\\
|\beta_{L,1}|-|\beta_{L,2}|, & |\beta_{L,2}\beta_{R,1}|\geqslant1\,\,\text{and}\,\,|\beta_{L,1}\beta_{R,2}|\leqslant1,\\
|\beta_{R,1}|-|\beta_{R,2}|, & |\beta_{L,1}\beta_{R,2}|\geqslant1\,\,\text{and}\,\,|\beta_{L,2}\beta_{R,1}|\leqslant1.
\end{cases}\label{eqn:cases}
\end{equation}
Note that $\zeta(\beta_{\alpha,j})$ is a function of $\lambda$ since $\beta_{\alpha,j}$ are functions of $\lambda$ through Eq.~(\ref{eq:char}). Therefore, we can solve Eq.~(\ref{eq:zeta}) as an equation of $\lambda$, and find the $\lambda$ eigenspectrum. From the $\lambda$ eigenspectrum we can obtain $\beta_{\alpha,j}$ by Eq.~(\ref{eq:char}). The $\beta_{\alpha,j}$ trajectories are the generalized Brillouin zones, which play a key role in the non-Hermitian bulk-boundary correspondence. The obtained generalized Brillouin zones are shown in
Fig.~\ref{fig:figS1}. To double check the validity of Eq.~(\ref{eq:zeta}) as the equation of generalized Brillouin zone, we also numerically diagonalize $U$ for finite-size systems, and then obtain the corresponding $\beta_{\alpha,j}$ by Eq.~(\ref{eq:char}). The generalized Brillouin zones obtained in this way are consistent with those obtained from Eq.~(\ref{eq:zeta}) [see Fig.~\ref{fig:figS1}].

\begin{figure*}
\includegraphics[width=0.4\textwidth]{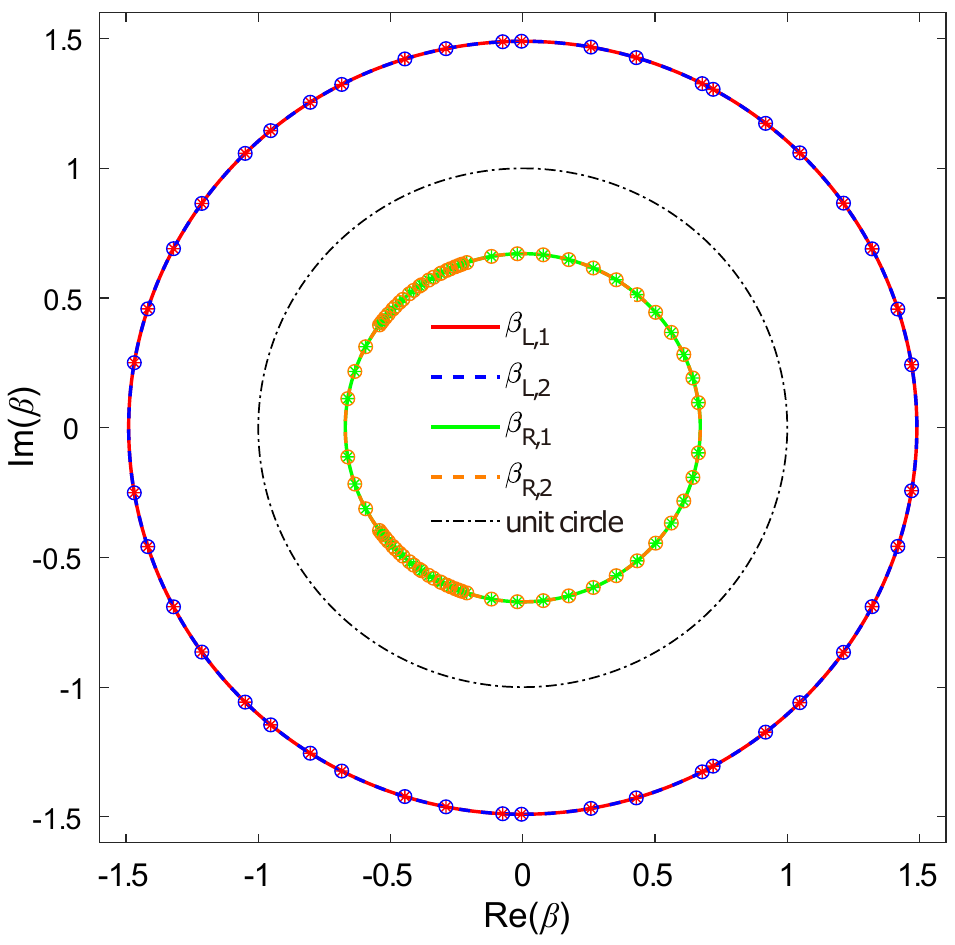}
\caption{{\bf Numerical check of the $\zeta$ function.} Generalized Brillouin zones calculated from Eq.~(\ref{eq:zeta}) (solid and dashed lines), compared to those obtained from the numerically calculated eigenenergy spectrum (circles and asterisks). In the latter approach, we numerically calculate the $\lambda$ eigenspectrum by diagonalizing $U$ of a finite size system, and then obtain $\beta_{\alpha,j}$ from Eq.~(\ref{eq:char}).  The two approaches lead to consistent results, confirming the validity of Eq.~(\ref{eq:zeta}) as the equation of generalized Brillouin zone. In this figure, we take the same parameters as those in Fig.~4b of the main text.
}
\label{fig:figS1}
\end{figure*}

Finally, we note that Eq.~(\ref{eqn:cases}) is helpful in identifying generalized Brillouin zones of the two bulks from numerically calculated eigenspectrum. Specifically, eigenstates of $U$ belonging to the first and second cases are associated with the generalized Brillouin zones of both bulks; whereas those belonging to the third (fourth) case are associated only with the generalized Brillouin zone of the left (right) bulk.

Based on the generalized Brillouin zones, we then calculate the non-Bloch topological invariants $\tilde{\nu}_\epsilon$ ($\epsilon=0,\pi$) defined in the main text. The results precisely match the topological edge modes with $\epsilon=0,\pi$ respectively, which embodies the non-Hermitian bulk-boundary correspondence.

\subsection{An alternative approach to calculate topological invariants:  Non-Bloch winding numbers of different time frames}

\begin{figure*}
\includegraphics[width=0.5\textwidth]{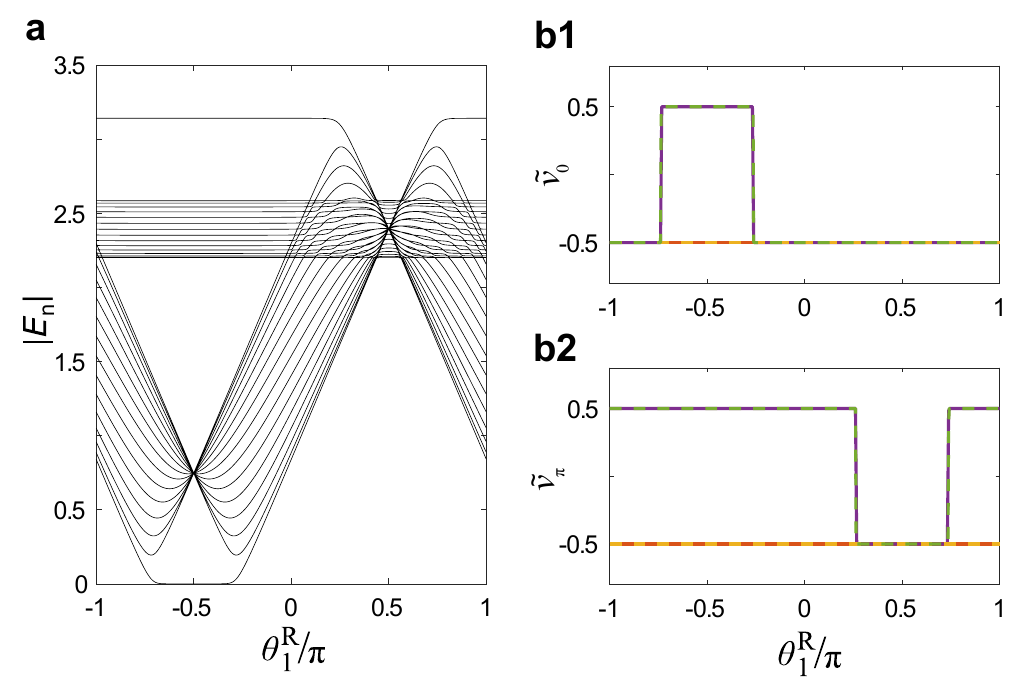}
\caption{{\bf Calculation of the non-Bloch topological invariants.}
{\bf a} The absolute values of the quasienergy spectrum as a function of $\theta_1^R$. The parameters are the same as those in Fig.~4a of the main text, where the quasienergy spectrum is completely real. {\bf b1, b2} Non-Bloch topological invariants of the left (red solid line) and right bulks (purple solid line) using periodized Floquet operators, as outlined in the main text. We also show non-Bloch topological invariants calculated using the two different time frames, for the left (yellow dashed line) and right bulks (green dashed line).
}
\label{fig:figS2}
\end{figure*}

\begin{figure*}
\includegraphics[width=\textwidth]{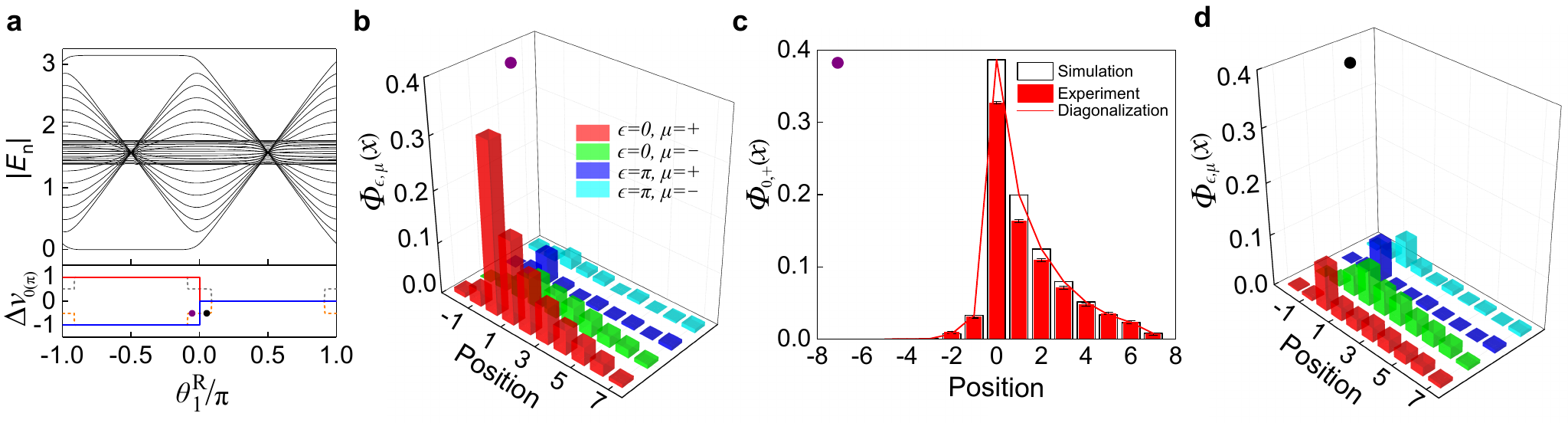}
\caption{{\bf Non-Hermitian bulk-boundary correspondence in the alternative time frame given by $U'$.}
{\bf a} Quasi-energy spectrum (black), and winding-number differences for the zero- (red) and $\pi$-modes (blue) between the two bulks with the parameters: $\theta_1^L=0.5625 \pi$, $\theta_2^R=0$, $\theta_2^L= \pi$, and $\gamma=0.2746$. The purple dot with $\theta_1^R=-0.0667 \pi$ corresponds to the parameter used in {\bf b}, where the system possesses both zero- and $\pi$-mode edge states. The black dot with $ \theta_1^R=0.0667 \pi$ corresponds to the parameter used in {\bf d}, where there is no edge state.
{\bf b} Experimentally measured $\varPhi_{\epsilon,\mu}(x)$ after the seventh step with the initial state $|0\rangle \otimes |+\rangle$.
{\bf c} Comparison between experimentally-measured and numerically-calculated $\varPhi_{0,+}(x)$ , as well as the scaled norms of the corresponding edge state after the seventh step. Topological edge states are numerically calculated for a domain-wall system with $N=15$, whose norms are scaled to fit the central peak of the corresponding $\varPhi_{0,+}(x)$.
{\bf d} Experimentally measured $\varPhi_{\epsilon,\mu}(x)$ after the seventh step with the same initial state as {\bf b}.
}
\label{fig:figS3}
\end{figure*}

\begin{figure*}
\includegraphics[width=0.5\textwidth]{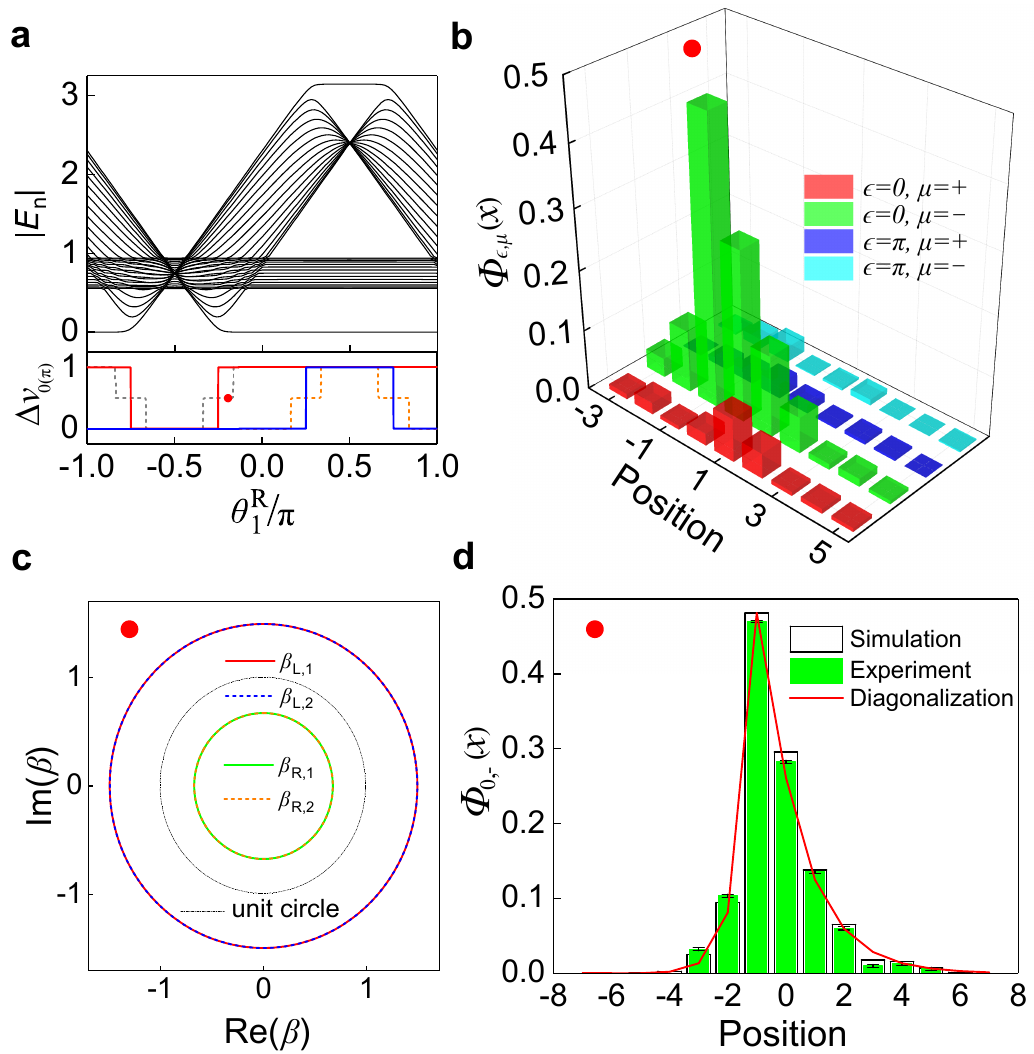}
\caption{{\bf Non-Bloch bulk-boundary correspondence for edge states with $\epsilon=0$.}  {\bf a} Quasi-energy spectrum (black), and winding-number differences $\Delta \tilde{\nu}_{0}$ (red) and  $\Delta \tilde{\nu}_\pi$ (blue) between the two bulks with the parameters: $\theta_1^L=-0.5625 \pi$, $\theta_2^R= 0.25\pi$, $\theta_2^L=0.75 \pi$, and $\gamma=0.2746$. Color and line shapes for winding numbers are the same as in Fig.~4 of the main text. The red dot with $\theta_1^R=-0.18 \pi$ corresponds to the coin parameter used in {\bf b, d}.
{\bf b} Experimentally measured $\varPhi_{\epsilon,\mu}(x)$ after the seventh step with the initial state $|0\rangle \otimes |-\rangle$.
{\bf c} Generalized Brillouin zones on the complex plane.
{\bf d} Comparison between experimentally-measured and numerically-calculated $\varPhi_{0,-}(x)$, as well as the scaled norms of the topological edge state after the seventh step.
}
\label{fig:figS4}
\end{figure*}

In this section, we provide an alternative approach to calculate the non-Bloch topological invariants. The idea is to calculate the non-Bloch winding numbers in different time frames of the Floquet sequence. In the Hermitian case, this approach has been adopted in Ref. \cite{AO13} to calculate the Bloch topological invariants.  The obtained topological invariants $\tilde{\nu}_0$ and $\tilde{\nu}_\pi$ below are the same as those calculated from the periodized Floquet operators $\overline{U}_\epsilon$.

First, we demonstrate how to calculate winding numbers in the time frame defined by $U$. The Fourier component of $U$ in the two bulks can be written as
\begin{equation}
U^{\alpha}(k)=d_{0}^{\alpha}\sigma_0-id_{1}^{\alpha}\sigma_{x}-id_{2}^{\alpha}\sigma_{y}-id_{3}^{\alpha}\sigma_{z},
\end{equation}
where
\begin{align}
&d_{0}^{\alpha}=-\cosh\gamma\sin\theta_{1}^{\alpha}\sin\theta_{2}^{\alpha}+\cosh\gamma\cos k\cos\theta_{1}^{\alpha}\cos\theta_{2}^{\alpha}+i\sinh\gamma\cos\theta_{1}^{\alpha}\sin k,\nonumber\\
&d_{1}=0,\nonumber\\
&d_{2}^{\alpha}=\cosh\gamma\cos\theta_{1}^{\alpha}\sin\theta_{2}^{\alpha}+\cos k\cosh\gamma\cos\theta_{2}^{\alpha}\sin\theta_{1}^{\alpha}+i\sin k\sinh\gamma\sin\theta_{1}^{\alpha},\nonumber\\
&d_{3}^{\alpha}=-\sin k\cosh\gamma\cos\theta_{2}^{\alpha}+i\cos k\sinh\gamma.
\end{align}
Here $\sigma_{x,y,z}$ are the standard Pauli matrices and $\sigma_0$ is the $2$-by-$2$ identity matrix. Near $k=0$, $d_3^\alpha$ resembles the $\sin k+i\gamma/2$ term appearing in the non-Hermitian Su-Schrieffer-Heeger model with non-Hermitian skin effect~\cite{Wang1}, which provides an intuitive understanding for the skin effect of $U^\alpha$.

To calculate the Bloch winding numbers, we follow the standard practice and apply a unitary transformation $V=\exp(i\frac{\pi}{4}\sigma_y)$ to $U^\alpha(k)$ such that
\begin{equation}
W^\alpha(k)=VU^\alpha(k) V^\dagger=d_{0}^{\alpha}I-i(-d_{3}^{\alpha})\sigma_{x}-id_{2}^{\alpha}\sigma_{y}-id_{1}^{\alpha}\sigma_{z}.
\end{equation}
The Bloch winding number is then defined through the generalized Zak phase
\begin{align}
&\nu^\alpha=\frac{\phi_{\rm Zak}^\alpha}{\pi},\\
&\phi_{\rm Zak}^\alpha=-\int_{-\pi}^{\pi}dk \frac{\langle\chi^\alpha_{k}|i\partial_{k}|\psi^\alpha_{k}\rangle}{\langle\chi^\alpha_k|\psi^\alpha_k\rangle},
\end{align}
where $|\psi^\alpha_k\rangle$ and $|\chi^\alpha_k\rangle$ are the right and left eigenstates of $W^\alpha(k)$ with
\begin{align}
&W^{\alpha}(k)|\psi_{k}^{\alpha}\rangle=E_{k}|\psi_{k}^{\alpha}\rangle,\\
&W^{\alpha\dagger}(k)|\chi_{k}^{\alpha}\rangle=E_{k}^{*}|\chi_{k}^{\alpha}\rangle,\\
&E_{k}^{\alpha}=d_{0}^{\alpha}-i\sqrt{(d_{1}^{\alpha})^{2}+(d_{2}^{\alpha})^{2}+(d_{3}^{\alpha})^{2}}.
\end{align}
From the above equations, we have
\begin{equation}
\nu^{\alpha}=\frac{1}{2\pi}\int dk\frac{-d_{3}^{\alpha}\frac{\partial d_{2}^{\alpha}}{\partial k}+d_{2}^{\alpha}\frac{\partial
d_{3}^{\alpha}}{\partial k}}{(d_{3}^{\alpha})^{2}+(d_{2}^{\alpha})^{2}}.
\label{eq:windingnumber}
\end{equation}

In contrast, under the non-Hermitian skin effect, bulk states become localized, therefore we need the non-Bloch winding numbers calculated along the generalized Brillouin zones. From the spatial-mode function $\beta_{\alpha,j}$, we define
\begin{align}
\beta_{\alpha,j}=|\beta_{\alpha,j}(p^\alpha_{j})|e^{ip^\alpha_{j}},
\end{align}
where $p^\alpha_{j}$ can be identified as the modified quasi-momentum in the $j$-th generalized Brillouin zone of the corresponding bulk.

In practice, it is sufficient to replace $e^{ik}$ with $\beta_{\alpha,j}$ in Eq.~(\ref{eq:windingnumber}), such that
\begin{equation}
\tilde{\nu}^{\alpha}=\frac{1}{2\pi}\oint dp_{j}^{\alpha}\frac{-\tilde{d}_{3,j}^{\alpha}\frac{\partial \tilde{d}_{2,j}^{\alpha}}{\partial p_{j}^{\alpha}}+\tilde{d}_{2,j}^{\alpha}\frac{\partial \tilde{d}_{3,j}^{\alpha}}{\partial p_{j}^{\alpha}}}{(\tilde{d}_{3,j}^{\alpha})^{2}+(\tilde{d}_{2,j}^{\alpha})^{2}},
\label{eq:nonblochnu}
\end{equation}
where
\begin{align}
&\tilde{d}_{2,j}^{\alpha}=\cosh\gamma\cos\theta_{1}^{\alpha}\sin\theta_{2}^{\alpha}+\cos\left(p_{j}^{\alpha}-i\ln|\beta_{\alpha,j}(p_{j}^{\alpha})|\right)\cosh\gamma\cos\theta_{2}^{\alpha}\sin\theta_{1}^{\alpha}+i\sin\left(p_{j}^{\alpha}-i\ln|\beta_{\alpha,j}(p_{j}^{\alpha})|\right)\sinh\gamma\sin\theta_{1}^{\alpha}\\
&\tilde{d}_{3,j}^{\alpha}=-\sin\left(p_{j}^{\alpha}-i\ln|\beta_{\alpha,j}(p_{j}^{\alpha})|\right)\cosh\gamma\cos\theta_{2}^{\alpha}+i\cos\left(p_{j}^{\alpha}-i\ln|\beta_{\alpha,j}(p_{j}^{\alpha})|\right)\sinh\gamma
\end{align}

The integration in Eq.~(\ref{eq:nonblochnu}) is over the $j$-th generalized Brillouin zone.
However, we have numerically checked that $\tilde{\nu}^\alpha$ calculated along different generalized Brillouin zones of a given bulk are the same. We therefore drop the index $j$ on the left-hand side of Eq.~(\ref{eq:nonblochnu}).

Following the procedure above, both Bloch and non-Bloch winding numbers in an alternative time frame can be calculated with the Floquet operator
\begin{align}
U'=M^{\frac{1}{2}}GFM^{\frac{1}{2}}.
\end{align}
Denoting the non-Bloch winding numbers of the two bulks as $\tilde{\nu}^{\prime\alpha}$, we calculate non-Bloch topological invariants $\tilde{\nu}^\alpha_0$ and $\tilde{\nu}^\alpha_\pi$ through
\begin{align}
\tilde{\nu}^\alpha_{0(\pi)}=\frac{\tilde{\nu}^\alpha\pm \tilde{\nu}^{\prime\alpha}}{2}.
\end{align}

These non-Bloch topological invariants are the same as those calculated using the periodized Floquet operators with branch cuts, and correctly predict the existence and number of topological edge states through the non-Hermitian bulk-boundary correspondence. In Fig.~\ref{fig:figS2}, we show a typical comparison between non-Bloch topological invariants calculated using the two methods. Furthermore, we have experimentally confirmed that adopting periodized Floquet operators associated with $U'$ would give the correct non-Bloch bulk-boundary correspondence in the alternative time frame. This is shown in Fig.~\ref{fig:figS3}.

\subsection{Non-Bloch bulk correspondence for edge states with $\epsilon=0$}

In the main text, we confirm non-Bloch bulk-boundary correspondence for the following three cases: i) only edge states with $\epsilon=\pi$ exist; ii) both types of edge states with $\epsilon=0$ and $\epsilon=\pi$ exist; iii) no edge state exists. For completeness, we also perform experiments using parameters under which only edge states with $\epsilon=0$ exist. This is shown in Fig.~\ref{fig:figS4}.

\end{widetext}

\end{document}